\documentclass[pra,aps]{revtex4}

\usepackage{graphicx}
\usepackage{dcolumn}
\usepackage{bm}
\usepackage{color} 

\begin{document}
\begin{titlepage}
\begin{center}
{\large \bf Erratum: Propagation of Second sound
in a superfluid Fermi gas in the unitary limit [Phys. Rev. A 80, 043613 (2009)]}\\
\vspace{5mm}
{ Emiko Arahata and Tetsuro Nikuni}
\vspace{10mm}
\end{center}


\setcounter{figure}{-1}
We made some errors in Sec.~IV.
The correct interaction parameter is $g_{\rm B}n_{\rm B}/k_{\rm B}T_{\rm c}=0.807$. 
The sound velocities plotted in Fig.~\ref{fig:becU2_U1} are normalized by the ``Fermi velocity" defined in terms of the density and mass of bosons, i.e.  $v_f=(6\pi^2 n_{\rm B})^{1/3}/2M$. 
Due to mishandling of raw data, the values given in Fig.~\ref{fig:becW2_W1} did not correspond to our calculated results.  The corrected version of the figure is given as Fig.~\ref{fig:bec} below. The qualitative behaviors of
$W_1$ and $W_2$ in the BEC limit are quite different from those in the unitary
limit, i.e. $W_2>W_1$ in most temperatures ($T>0.4T_{\rm c}$).
This change does not affect any
other result nor the conclusion of the paper.
\begin{figure}[htbp]
\centerline{\includegraphics[height=2.0in]{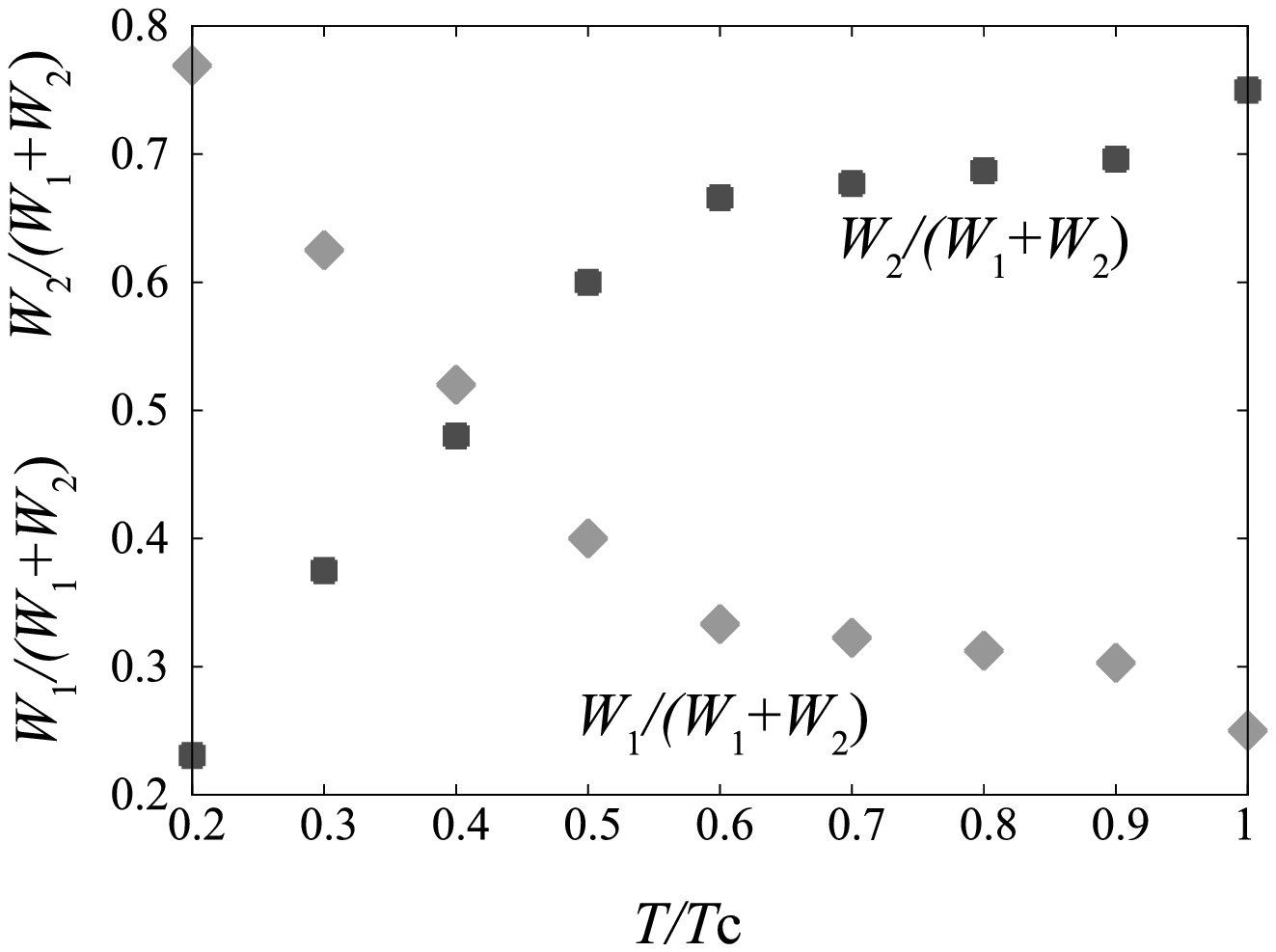}}
 \caption{$W_1/(W_1+W_2)$ and $W_2/(W_1+W_2)$ as a function of temperature.}
  \label{fig:bec}
\end{figure}

\end{titlepage}

\title{Propagation of Second sound
in a superfluid Fermi gas in the unitary limit}
\author{Emiko Arahata}

\author{Tetsuro Nikuni}%
\affiliation{%
Department physics, Faculty of science, Tokyo University of Science, \\
1-3 Kagurazaka, Shinjuku-ku, Tokyo 162-8601, Japan}%
\date{\today}
\begin{abstract}
We study sound propagation in a uniform superfluid gas of Fermi atoms in the unitary limit. 
The existence of normal and superfluid components leads to appearance of two sound modes in the collisional regime, referred to as first and second sound. The second sound is of particular interest as it is a clear signal of a superfluid component. Using Landau's two-fluid hydrodynamic theory, we calculate hydrodynamic sound velocities and these weights in the density response function. The latter is used to calculate the response to a sudden modification of the external potential generating pulse propagation. The amplitude of a pulse which is proportional to the weight in the response function, is calculated the basis of  the approach of Nozi$\rm{\grave{e}}$res and Schmitt-Rink (NSR) for the BCS-BEC crossover. We show that, in a superfluid Fermi gas at unitarity, the second sound pulse is excited with an appreciate amplitude by density perturbations.
\pacs{03.75.Kk, 03.75.Ss, 67.25.D-}\end{abstract}
\setcounter{page}{1}
\maketitle 
\section{Introduction}
Landau's two-fluid hydrodynamics describes
the finite temperature dynamics of all superfluids when collisions are sufficiently
strong to produce a state of local thermodynamic equilibrium \cite{Hydro_Landau}. Recent experiments have begun to observe sound propagation in trapped superfluid Fermi gases with a Feshbach resonance \cite{E_Kinast_PRL92,E_Bartenstein_PRL92,E_Joseph_PRL98}. At unitarity, the magnitude of the $s$-wave scattering length that characterizes the interactions
between fermions in different hyperfine states diverges ($|a_s|\to \infty$). Owing to the strong interaction close to unitarity, the dynamics of superfluid Fermi gases with a Feshbach resonance at finite temperatures are expected to be described by
Landau's two-fluid hydrodynamic equations \cite{T_Taylo_sound}. Two-fluid hydrodynamics predicts the existence of in-phase modes in which the superfluid and normal fluid components move together, as well as out-of-phase modes where the two components move against to each other. These two sound modes in the collisional limit are referred to as first and second sounds. Of greater interest is the out-of-phase second sound mode, since it is a clear signal of the existence of a superfluid component. Out-of-phase hydrodynamic modes in strongly-interacting Fermi superfluids have been discussed theoretically in the literature. The propagations of first and
second sound in a uniform superfluid at unitarity are discussed in Refs.~\cite{T_Heiselberg_Unitarysound,T_Fukushima_PRA75,T_Taylor_PRA77}. References \cite{T_Yan_sound,T_Taylor_PRA77,T_condmat07090698,T_condmat09050257} studied out-of-phase collective modes in trapped Fermi gases, which are more relevant to experiments. However, out-of-phase modes have not been observed experimentally so far.  
\par Experimentally, the sound wave in a highly elongated trapped gas can be excited by a sudden modification of a trapping potential using the focused laser beam. The resulting density perturbations propagate with a speed of sound. This technique was first used to probe Bogoliubov sound in a Bose-condensed gas \cite{E_Andrews_PRL79}. Observed sound velocity was in good agreement with theoretical predictions \cite{T_PRA78,T_PRA57}.  Analogous sound propagations have been discussed for a normal Bose gas \cite{T_Nikuni_}. Possibility of observing propagation of first and second sound pulses in a Bose-condensed gas was also briefly discussed in Ref.~\cite{T_Nikuni_}.  Sound propagation was also studied theoretically for a normal Fermi gas in Ref.~\cite{T_C}. More recently, the pulse technique was used to study sound propagation in a Fermi gas near a Feshbach resonance \cite{E_Joseph_PRL98}. In this experiment, first sound mode was observed but second sound mode was not observed. 
In principle, one should be able to probe two-fluid hydrodynamic sound modes using this technique. 
\par 
In the present paper, we discuss sound pulse propagation in a strongly interacting Fermi gas in the two-fluid hydrodynamic regime. In Ref.~\cite{T_Heiselberg_Unitarysound}, the first and second sound velocities in the BCS-BEC crossover for a uniform gas was estimated  theoretically. Reference~\cite{T_Heiselberg_Unitarysound} also argued that both sound modes can be excited and detected both as density and thermal waves, but the quantitative results were not presented. Reference \cite{T_condmat07090698} calculated the two-fluid density response spectrum in a uniform superfluid gas of Fermi atoms in the unitary limit, and showed that second sound is only weakly coupled into density response \cite{T_condmat07090698}. At first sight, this result seems to  imply that the second sound cannot be excited by a density perturbation.  In fact, it would not show up in Bragg scattering.   However, we will show that second sound can be still observed by a sudden modification of the external potential generating pulse propagation.
In this paper, we use Landau's two-fluid hydrodynamic equations to study pulse propagation in a unitary Fermi gas.  \par
In Sec. II, we discuss the linear response solutions of the Landau's
two-fluid hydrodynamic equations for uniform superfluid gases. We show that sound pulse propagation is described in terms of the density response function. The amplitudes of the first and second sound pulses are explicitly expressed in terms of the weights in the density response spectrum. In Sec. III, we use the Nozi$\rm{\grave{e}}$res and Schmitt-Rink (NSR) theory to calculate thermodynamic quantities, which are needed as inputs in our solutions of the two fluid equations. These results are used to calculate the temperature dependence of velocity and pulse amplitude of the second sound mode in Sec.~IV. We find that second sound has an appreciable weight in the propagation of density pulses. For composition, in Sec.~V, we calculate the temperature dependence of velocity and amplitude of the second sound pulse in the BEC limit using Hatrree-Fock-Bogoliubov-Popov (HFB-Popov) approximation.

\section{Linear Response solution of Landau's two-fluid equations }
In this section, we present a solution of Landau's two fluid hydrodynamic  equation in the presence of external perturbation within the liner response theory. We review normal mode solutions.   
The Landau two-fluid hydrodynamic equations in a uniform superfluid are given by \cite{B_Pethich_Smith,B_GNZ}
\begin{eqnarray}
m \frac{\partial \mathbf{j}}{\partial t}=-\nabla P,\label{Leq_j}\\
\frac{\partial n}{\partial t} +\nabla \cdot \mathbf{j}=0, \label{Leq1} \\
\frac{\partial s}{\partial t}+\nabla \cdot(s\mathbf{v}_n)=0 \label{Leq2},\\
m\frac{\partial \mathbf{v}_s}{\partial t}=-\nabla\mu.\label{Leq_vs}
\end{eqnarray}
The total mass current
\begin{eqnarray}
m\mathbf{j}=\rho_s\mathbf{v}_s+\rho_n\mathbf{v}_n,
\end{eqnarray} 
is given in terms of the superfluid and normal fluid velocities
$\mathbf{v}_s$ and $\mathbf{v}_n$, as well as the superfluid and normal fluid densities,
$\rho_s$ and $\rho_n$. The sum of the superfluid and normal fluid densities
gives the total mass density, $mn=\rho=\rho_s+\rho_n$. The continuity
equation in Eq.~(\ref{Leq1}) expresses mass conservation and is always valid. Equation~(\ref{Leq2}) assumes that the entropy of the
fluid is carried by the normal fluid and is conserved. These
equations describe reversible flow without any dissipation
arising from transport coefficients \cite{B_Pethich_Smith,B_GNZ}.
We now consider the linearized Landau equations for a uniform superfluid. The linearized continuity and entropy conservation equations given by Eqs.~(\ref{Leq1}) and (\ref{Leq2}) are 
\begin{eqnarray}
m \frac{\partial \delta n}{\partial t} +\nabla \cdot \left(\rho_{s0}v_s+\rho_{n0}v_n\right)=0, \label{Leq1p} \\
\frac{\partial \delta s}{\partial t}+\nabla \cdot(s_0 v_n)=0 \label{Leq2p}.
\end{eqnarray}
Taking time derivative of Eqs.~(\ref{Leq1p}) and (\ref{Leq2p}), and combining them with Eqs.~(\ref{Leq_j}) and (\ref{Leq_vs}) (in linearized forms) in conjunction with the thermodynamic identity $n_0\mu=S_0 \delta T+\delta P$, we arrive at a closed set of equations in terms of the variables $\delta \rho$ and $\delta s$. Inserting the normal mode plane-wave solution $\delta \rho, \delta s \propto e^{i(\mathbf{q}\cdot \mathbf{r}-\omega t)}$, one finds $\omega^2=u^2 q^2$, where $u$ is given by 
\begin{eqnarray}
u^2=\frac{C_s^2+C_2^2}{2}\pm\sqrt{\left(\frac{C_s^2+C_2^2}{2}\right)^2-C_T^2C_2^2}. \label{u12}
\end{eqnarray}
The thermodynamic quantities entering are the adiabatic sound speed squared $C_s^2=\left(\frac{\partial P}{\partial \rho}\right)_{\bar{s}}$, the isothermal and the thermal sound speed squared $C_T^2=\left(\frac{\partial P}{\partial \rho}\right)_{T}$, $C_2^2=\frac{\rho_{s0}}{\rho_{n0}}\frac{T\bar{s}^2_0}{\bar{C_v}}$. The latter also acts as a
coupling or mixing term. The difference between the adiabatic
and isothermal sound speed squared can also be expressed
as $C_s^2-C_T^2=\left(\frac{\partial s}{\partial \rho}\right)_T^2\frac{\rho^2T}{c_v}$. Here, $\bar{s}=s/\rho$ the entropy per unit mass, and $C_v=T\left(\frac{\partial \bar{s}}{\partial T}\right)_\rho$ the specific heat per unit mass. \par 
We now consider the two-fluid hydrodynamics in the presence of an external time-dependent potential $\delta U(\mathbf{r},t)$. In this case, the equations for $\mathbf{j}$ and $\mathbf{v}_s$ become
\begin{eqnarray}
m\frac{\partial \mathbf{j}}{\partial t}=-\nabla P-n\delta U,\\
m\frac{\partial \mathbf{v_s}}{\partial t}=-\nabla (\mu+\delta U),
\end{eqnarray}
Within the linear response theory, the general solution for the density fluctuation $\delta n$ can be written in terms of the density-density response function as 
\begin{eqnarray}
\delta n(\mathbf{r},t)=\int \frac{ d\mathbf{q}}{(2\pi)^3} \int \frac{d\omega}{2\pi} \chi_{nn}(\mathbf{q},\omega)\delta U(\mathbf{q},\omega)e^{i\mathbf{q}\cdot \mathbf{r}-i\omega t},
\end{eqnarray}
where $\delta U(\mathbf{q}, \omega)=\int d \mathbf{r}\int dt \delta U(\mathbf{r},t)e^{-i\mathbf{q}\cdot \mathbf{r}+i\omega t} $ is the Fourier transform of the external potential. The density response function for a uniform superfluid described by the Landau's two-fluid equation is given by \cite{B_GNZ}
\begin{eqnarray}
\chi_{nn}(\mathbf{q},\omega)=\frac{n_0 q^2}{m}\frac{\omega^2- v^2q^2}{(\omega^2-u_1^2 q^2)(\omega^2-u_2^2 q^2)},\label{response1}
\end{eqnarray}
where a new velocity $v$ is defined by
\begin{eqnarray}
v^2=\bar{s}^2_0\frac{\rho_{s0}}{\rho_{n0}}\frac{\partial T}{\partial \bar{s}}.\label{v2}
\end{eqnarray}
The two-fluid density response function in (\ref{response1}) was first derived for superfluid $^4$He by Ginzburg \cite{Hydro_Ginz} and Hohenberg and Martin \cite{T_Hohenberg_PRL12}. It was  first applied to weakly interacting superfluid Bose gases by Gay and Griffin \cite{T_Gay_sound}. In the case of the sound propagation experiment \cite{E_Kinast_PRL92,E_Bartenstein_PRL92,E_Joseph_PRL98}, a localized potential is applied at $t>0$, while it is turned off at $t=0$. This situation can be described as 
\begin{eqnarray}
\delta U(\mathbf{r},t)=\delta U(z)\theta(-t).\label{tita}
\end{eqnarray} 
Here we assume that the external potential is uniform in the $xy$ direction and is localized at $z\simeq 0$. In this case, the density fluctuations at ($t>0$) is given by 
\begin{eqnarray}
\delta n(z,t)=\frac{1}{2\pi^2}\int  dq\int d\omega \delta U(q)
\frac{\chi_{nn}^{\prime\prime}}{(w+i\eta)}e^{iqz-i\omega t} \ \ (t>0),\label{delta n3}
\end{eqnarray}
where $\chi_{nn}^{\prime\prime} (\mathbf{q},\omega)= {\rm{Im}}\chi_{nn}(\mathbf{q},\omega+i\eta)$. More explicitly, it is written as
\begin{eqnarray}&&{\rm{Im}}\chi_{nn}(\mathbf{q},\omega+i\eta)\nonumber\\
&&=\pi \frac{q^2}{m}Z_1\delta(\omega^2-u_1^2 q^2)+\pi \frac{q^2}{m}Z_2\delta(\omega^2-u_2^2 q^2),\label{kaip}
\end{eqnarray} 
where 
\begin{eqnarray}
Z_1=\frac{u_1^2-v^2}{u_1^2-u_2^2}, \ \ \
Z_2=-\frac{u_2^2-v^2}{u_1^2-u_2^2}=1-Z_1. \label{Z12}
\end{eqnarray}
Using Eqs.~(\ref{kaip}) and (\ref{W12}) in (\ref{delta n3}), we obtain 
\begin{eqnarray}
\delta n(z,t)=W_1\left[\delta U(z-u_1t)+\delta U(z+u_1t)\right]
+W_2\left[\delta U(z-u_2t)+\delta U(z+u_2t)\label{delta n_u}\right].\label{den}
\end{eqnarray}
where 
\begin{eqnarray}
W_1=\frac{n_0}{2mu_1^2}Z_1=\frac{n_0}{2mu_1^2}\frac{u_1^2-v^2}{u_1^2-u_2^2}, \ \ \
W_2=\frac{n_0}{2mu_2^2}Z_2=-\frac{n_0}{2mu_2^2}\frac{u_2^2-v^2}{u_1^2-u_2^2}. \label{W12}
\end{eqnarray}
\par 
The expression Eq.~(\ref{delta n_u}) describes propagation of sound pluses with the speeds $u_1$ and $u_2$ with the amplitudes $W_1$ and $W_2$. We note that the above general result applies to dissipationless dynamics of all superfluid in the collisional hydrodynamics regime. However, details are quite different for different systems. For example, in superfluid $^4$He, we have $u_2\simeq v$ and hence $Z_2\simeq 0$. In this case only first sound can be excited by the density perturbation. In contrast,
second sound can have an appreciable weight in the density response function in superfluid Bose gases at finite temperatures. The main purpose of the present paper is to show that the second sound can be excited by the density perturbation of the form (\ref{tita}) in a superfluid Fermi gas at unitarity. \par
The density response spectrum $\chi_{nn}^{\prime\prime} (\mathbf{q},\omega)$ in a superfluid Fermi gas at unitarity was calculated in Ref.~\cite{T_condmat07090698}. The result of Ref.~\cite{T_condmat07090698} showed that the weight of first sound is everywhere much larger than second sound. Second sound is only weakly coupled into the density response function ($Z_2\simeq0.05$ is the maximum at $T\simeq0.9 T_{\rm{c}}$ ). However, as shown in Eq.~(\ref{W12}), the pulse amplitude $W_i$ involves an extra factor ($1/u_i^2$) which arises due to the pulse perturbation of the form (\ref{tita}). Since in general $u_2 < u_1$, the second sound pulse amplitude $W_2$ is relatively amplified, and can be much larger than the weight in $\chi^{\prime\prime}(\mathbf{q},\omega)$.  
In the following sections, we explicitly calculate $u_1$, $u_2$, and $v$ for a superfluid Fermi gas using the microscopic theory.
\section{Thermodynamic functions}
The explicit calculation of the weights $W_1$ and $W_2$ in (\ref{W12}) requires thermodynamics quantities, such as $\rho_{s0}, \ \rho_{n0}, \ (\partial P/\partial \rho )_T,\  \bar{s}$, and so on.
In this section, we discuss the approximations used to evaluate these quantities. The calculation  is based on the Leggett mean-field BCS model of the BCS-BEC
crossover, extended to include the effects of pairing fluctuations
associated with the dynamics of the bound states using
the approach of Nozi$\grave{e}$res and Schmitt-Rink (NSR) \cite{T_NSR,T_Engelbrecht_PRB55,T_Ohashi_PRA}. The NSR approximation has also been used to calculate the thermodynamic properties in the BCS-BEC crossover at both $T$
=0 and finite temperatures \cite{T_NSR,T_Taylor_PRA74}. In the NSR
theory, the superfluid order parameter $\Delta$ and chemical potential $\mu$ are determined from the coupled equations \cite{T_Fukushima_PRA75}. 
\begin{eqnarray}
1&=&-\frac{4\pi a_s}{m}\sum_{p}\left(\frac{1}{2E_p}\tanh\frac{\beta E_p}{2}-\frac{1}{2\epsilon_p}\right),\\
N&=&\sum_p\left(1-\frac{\xi_p}{E_p}\tanh\frac{\beta E_p}{2}\right)-\frac{1}{2\beta}\frac{\partial }{\partial \mu}\sum_{q,\nu_n}\ln\det\left \{1-\frac{4\pi a_s}{m}\left[\Xi(\mathbf{q},i\nu_n)+\frac{1}{2\epsilon_p}\right]\right\},\label{N}
\end{eqnarray}
where the single-particle quasiparticle energies are given by $E_p=\sqrt{\xi_p^2+\Delta^2}$ with $\xi_p\equiv\epsilon_p-\mu, \ \ \ \epsilon_p\equiv\frac{\hbar^2 p^2}{2m}$. The two-body $s$-wave
scattering length is denoted as $a_s$, and $\nu_n$ is the bosonic Matsubara frequency. 
The second term in Eq.~(\ref{N}) describes contribution
from bosonic collective pair fluctuations \cite{T_Fukushima_PRA75}, where the expression for $\Xi$ is given by in Appendix \ref{C_1}.
The key function of interest in this paper is the thermodynamic
potential, defined by
  \begin{eqnarray}
 \Omega&=&-|\Delta|^2\frac{m}{4\pi a_s}-\frac{1}{\beta}\sum_\mathbf{k}\mathrm{tr}\ln\left[-\mathbf{G}_0^{-1}(k)\right]+\frac{1}{2\beta}\sum_{\mathbf{q},\nu_n}\ln\det \left[1+\frac{4\pi a_s}{m}\Xi(\mathbf{q},i\nu_n)\right], \\
 &&\mathbf{G}_0^{-1}(k)\equiv
 \left(\begin{array}{cc}i\hbar \omega_m- \xi_k& \Delta \\ \Delta^\ast & i\hbar \omega_m+ \xi_k\end{array}\right)  \delta_{k,k^\prime}\delta_{m,m^\prime}.
\end{eqnarray}
All thermodynamic quantities of interest can be calculated
once $\Omega$ is given. For example, we can then calculate pressure by using the relation $P=-\frac{\Omega}{V}$ first. We can obtain pressure terms $(\partial P/\partial \rho)_T$ using numerical differentiation. We use the relation $C_s^2-C_T^2=\left(\frac{\partial s}{\partial \rho}\right)_T^2\frac{\rho^2T}{c_v}$ to calculate $C_s^2$ because it is difficult to calculate $(\partial P/\partial \rho)_s$ numerically. Due to the difficulty in numerical differentiation of thermodynamic quantities at low temperature $T/T_{\rm{c}}<0.2$, we calculate $u_2$, and $W_2$ only for $T/T_{\rm{c}}>0.2$. \par
The superfluid density $\rho_s$ can also be obtained from the
thermodynamic potential \cite{T_NSR,T_Taylor_PRA74}. 
The normal fluid density $\rho_n$ associated
with fermionic and bosonic degrees of freedom is given by
the sum of their contributions:
\begin{eqnarray}
\rho_n=-\frac{2}{3m}\sum_p p^2\frac{\partial f_{\rm{FD}}(E_p)}{\partial E_p}-\frac{2m}{\beta} \frac{\partial }{\partial Q_z}\sum_{q,\nu_n} \frac{U}{\eta(\mathbf{q},i \nu_n)}\Bigg\{\left[1+U\Pi^0_{11}(\mathbf{q},i\nu_n)\right]\frac{\partial \Pi^0_{22}(\mathbf{q},i\nu_n)}{\partial Q_z}\nonumber \\
+\left[1+U\Pi^0_{22}(\mathbf{q},i\nu_n)\right]\frac{\partial \Pi^0_{11}(\mathbf{q},i\nu_n)}{\partial Q_z}-2U\Pi^0_{12}(\mathbf{q},i\nu_n)\frac{\partial \Pi^0_{12}(\mathbf{q},i\nu_m)}{\partial Q_z}\Bigg\}_{Q_z \to 0}.\label{rho_n}
\end{eqnarray}
Here, $f_{\rm{FD}}(E)=1/(e^{\beta E}+1)$ is the Fermi-Dirac distribution function, $\eta(\mathbf{q},i \nu_n)=\det \left[1+U\Xi(\mathbf{q},i\nu_n)\right]$, $U=\frac{4\pi a}{m}$, and a supercurrent flows in the $z$ direction with the
superfluid velocity is $v_s=Q_z/2m$.
We can obtain the superfluid density $\rho_s$ from the relation $\rho_s=\rho-\rho_n$. \par
\begin{figure}[htbp]
\centerline{\includegraphics[height=2.0in]{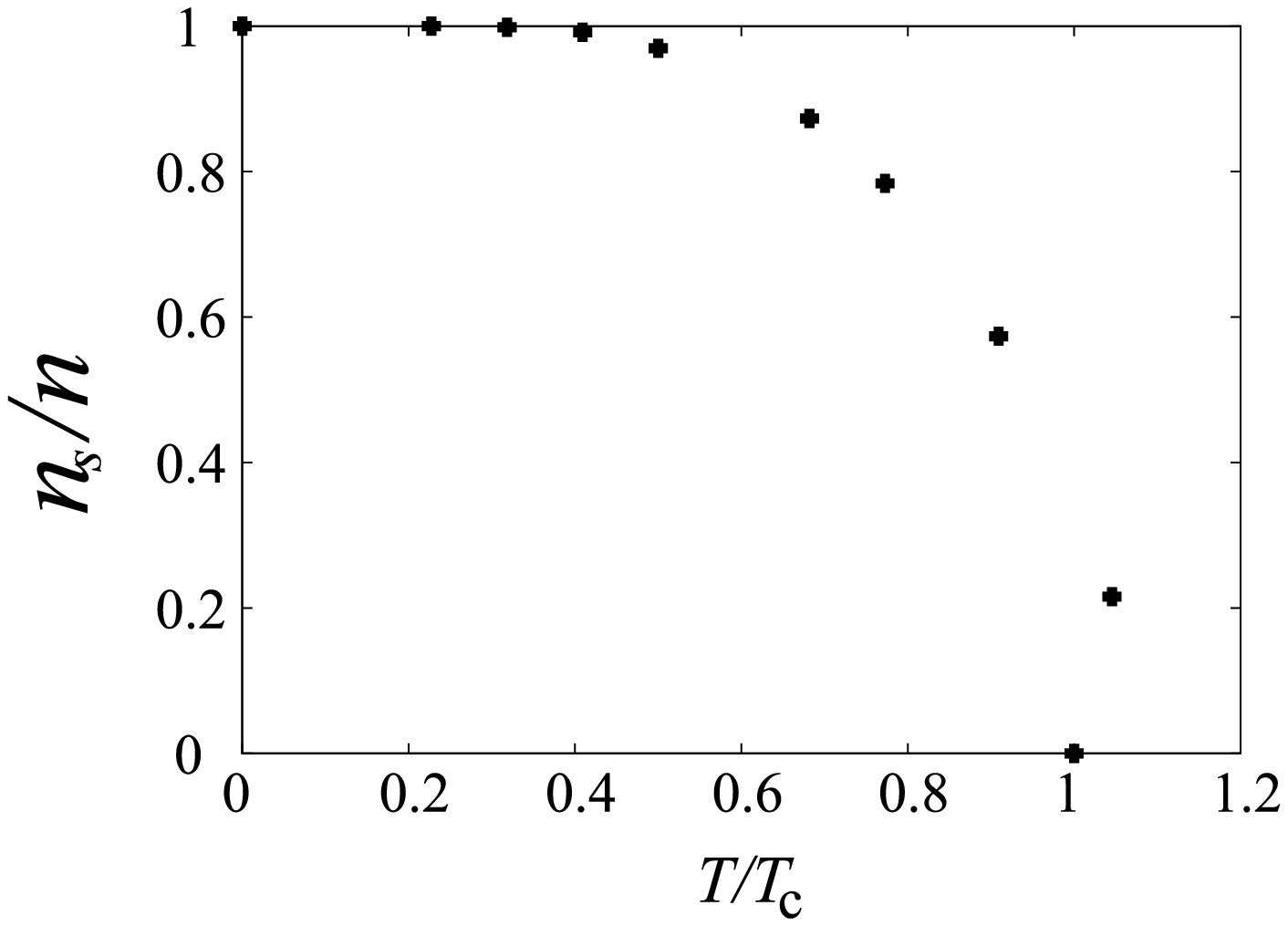}}
 \caption{Superfluid density fraction in a uniform
Fermi gas at unitarity as a function of temperature.} 
  \label{fig:rhos}
\end{figure}

Now we present numerical results for the superfluid
density $\rho_s$, starting from the expression for $\rho_n$
given in Eq.~(\ref{rho_n}).
Our calculation procedure closely follows that summarized in Refs.~\cite{T_Fukushima_PRA75}. Fig.~\ref{fig:rhos} shows the calculated superfluid density $\rho_s$ at unitarity $a_s\to\infty$. 
The NSR theory does have a problem near $T_{\rm{c}}$ near unitarity and on the BEC side of the crossover as a result of only considering Gaussian fluctuations.
However, we consider it the best available theory for
the thermodynamic variables at finite temperatures in the BCS-BEC crossover at the present time.
\section{Sound Propagation}
\begin{figure}[htbp]
\centerline{\includegraphics[height=2.0in]{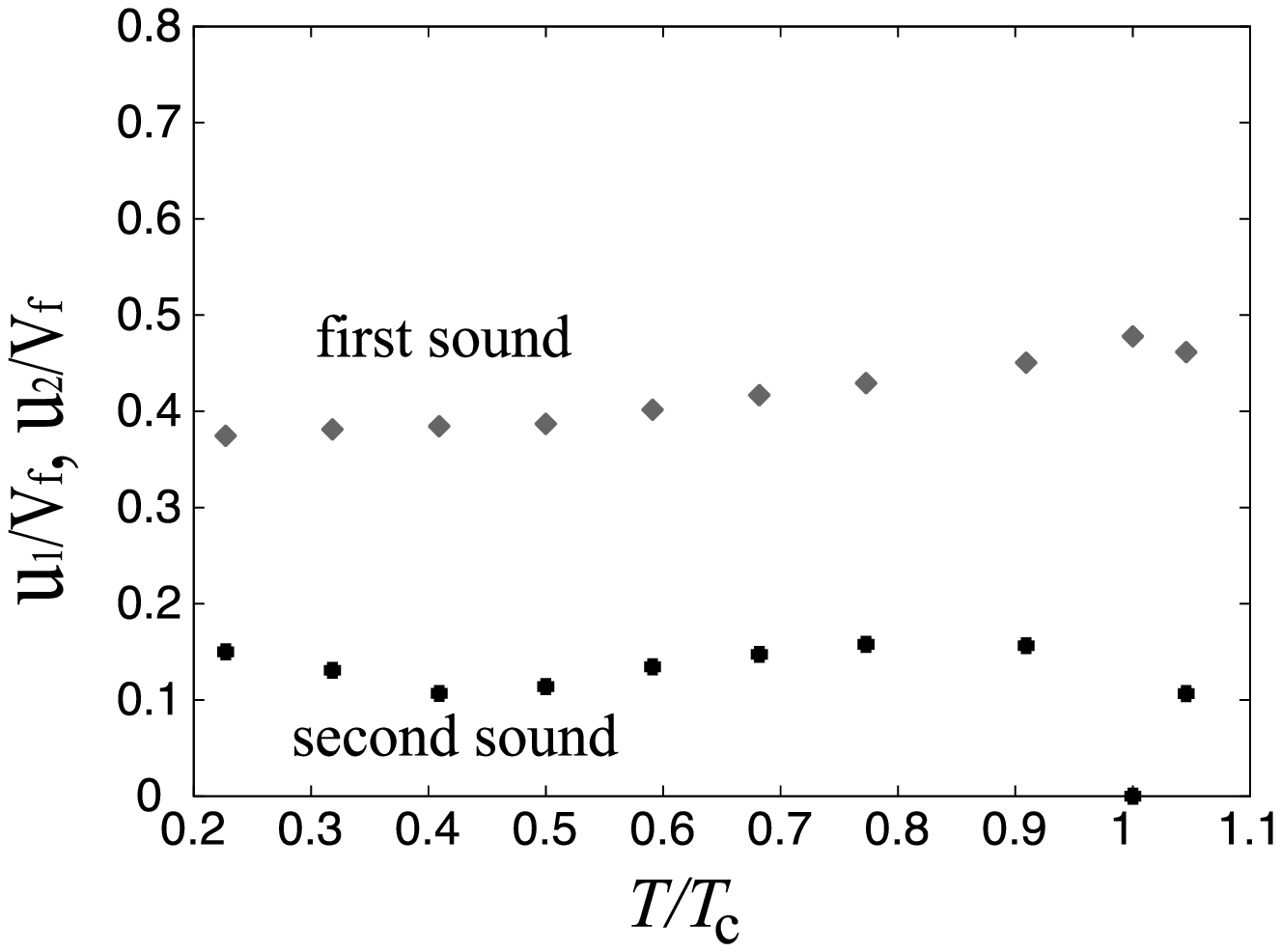}}
 \caption{The sound velocities in a uniform
Fermi gas at unitarity as a function of temperature.  $v_f=\hbar k_f/m$. $k_f$ is Fermi wave number} 
  \label{fig:u2}
\end{figure}

Using the thermodynamic quantities calculated in the previous section, we discuss first and second sound propagation. Since we are interested in the unitary limit, we set $1/a_s=0$.  
In Fig.~\ref{fig:u2}, we plot the sound velocities in a uniform
Fermi gas at unitarity as a function of temperature.
Near the critical temperature, the second sound velocity approaches zero. The second sound velocity has a broad maximum around $T\sim0.9T_{\rm{c}}$.
The NSR-type theories developed in Refs. \cite{T_Hu_PRA73,T_NSR,T_Taylor_PRA74,T_Fukushima_PRA75} only includes the contributions from the BCS Fermi excitations plus the bosonic pairing fluctuations. This leads a problem to calculate the velocity of second sound near $T_{\rm{c}}$.
\begin{figure}[htbp]
\centerline{\includegraphics[height=2.0in]{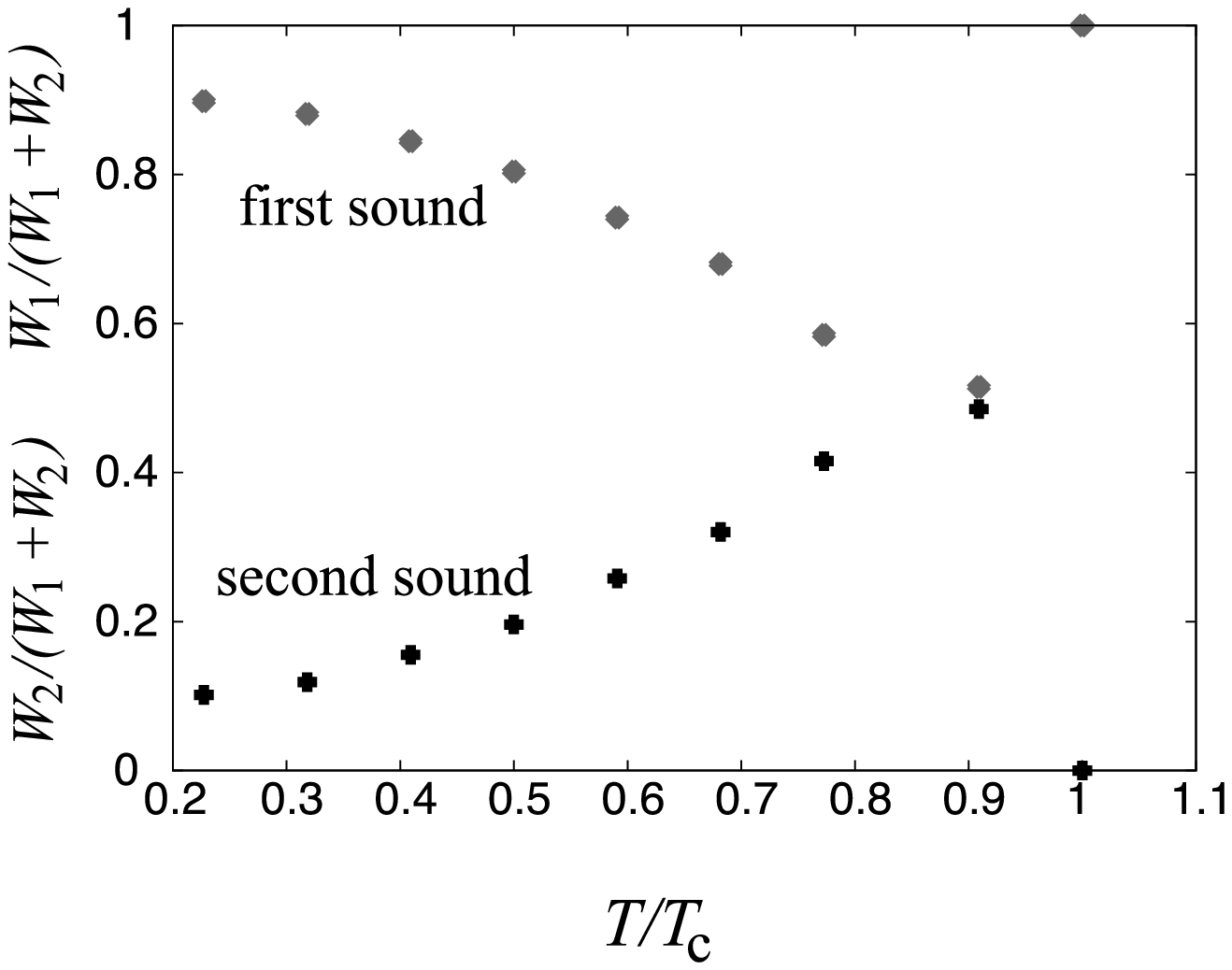}}
 \caption{The first sound amplitude $W_1/(W_1+W_2)$ (circle) and the second sound amplitude $W_2/(W_1+W_2)$ (diamond) as a function of temperature. } 
  \label{fig:W2}
\end{figure}
\par In Fig.~\ref{fig:W2}, we plot the temperature dependence of $W_1/(W_1+W_2)$ and $W_2/(W_1+W_2)$ obtained by the NSR-type Gaussian fluctuation theory discussed in the previous section.
One immediately sees that second sound pulse has an appreciable amplitude. The second sound amplitude decreases at low temperatures and increases with increasing temperature, before decreasing again
as $T_{\rm{c}}$ is approached. The second sound amplitude has a sharp maximum around $T\sim 0.9T_{\rm{c}}$.
\begin{figure}[htbp]
\centerline{\includegraphics[height=2.0in]{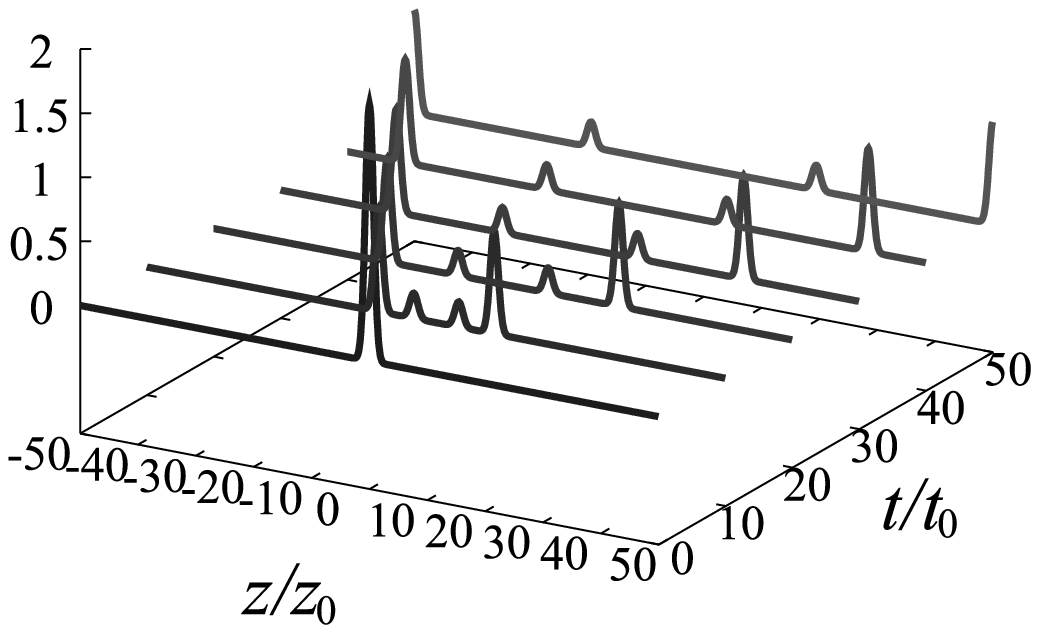}}
 \caption{The perturbed density profile $T=0.6T_{\rm{c}}$ for several
propagation times. $z_0=1/k_f$ and $t_0=1/(k_fu_1)$} 
  \label{fig:st}
\end{figure}
Figure \ref{fig:st} shows the perturbed density profile at $T=0.6T_{\rm{c}}$ for several
propagation times. Since $u_1>u_2$, the second sound pulse propagates slower than the first sound pulse. We clearly see that second sound is excited by density perturbations in the superfluid Fermi gas at unitarity. Our results show that both sound modes can be observed by a sudden modification of the external potential using a pulse wave. As discussed in Sec. II, although the weight $Z_2$ is very small in the unitary Fermi gas, the pulse amplitude $W_2$ is amplified by a factor of to because of the low velocity of the second sound.    
\begin{figure}[htbp]
\centerline{\includegraphics[height=2.0in]{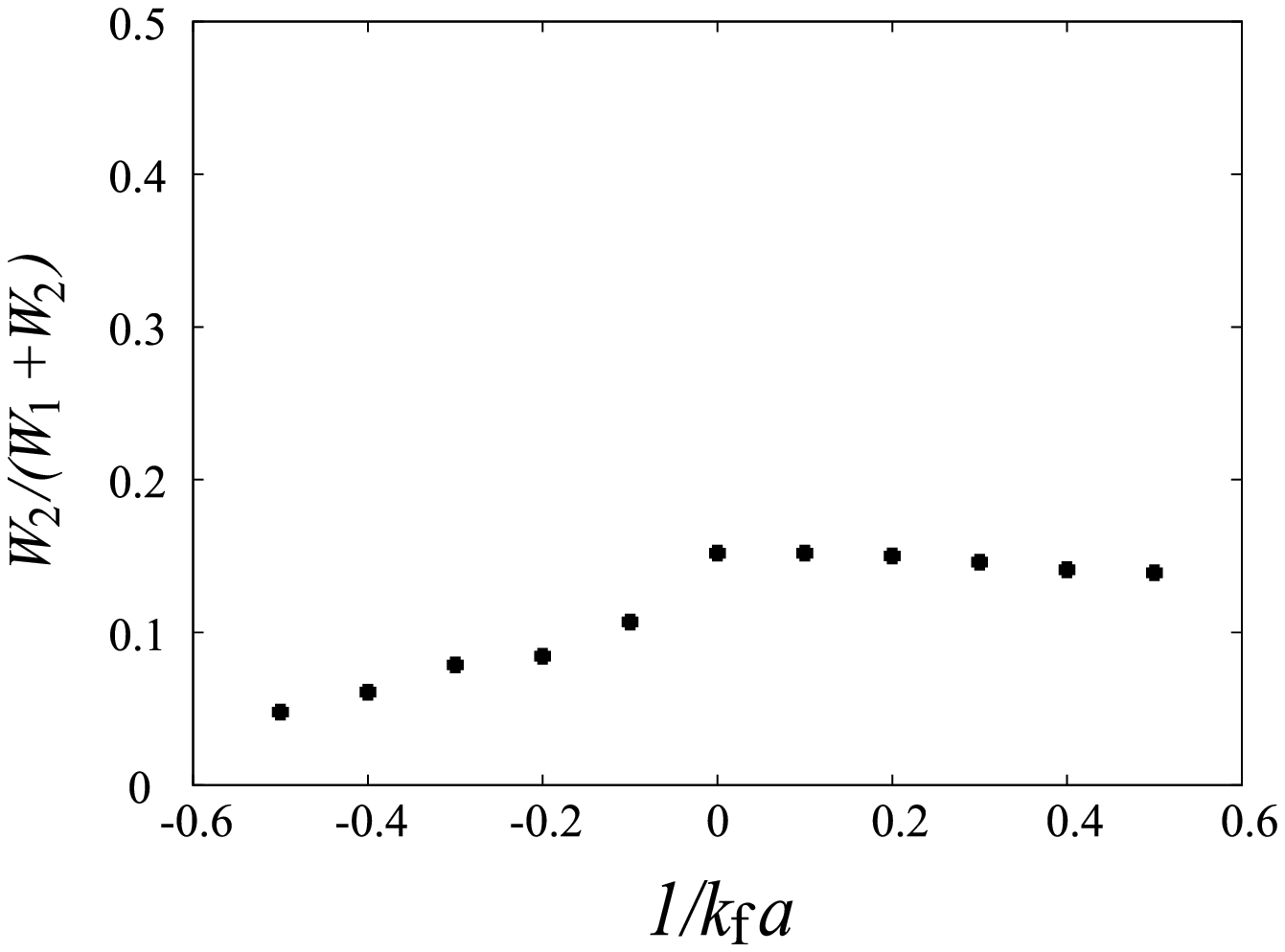}}
 \caption{the second sound amplitude $W_2/(W_1+W_2)$ as a function of $1/k_fa$.} 
  \label{fig:W2_closs}
\end{figure}
\par We now briefly discuss the effect of changing the scattering length. 
In Fig.~\ref{fig:W2_closs}, we plot $W_2/(W_1+W_2)$ as a function of $1/k_fa$ with fixing the temperature as $T/T_{\rm{c}}=0.6$, where $T_{\rm{c}}$ is the superfluid transition temperature at a given $1/k_f a$. We see that second sound pulse has an appreciable weight over a finite range in the crossover region. The second sound amplitude has a broad maximum around $\frac{1}{k_f a} \simeq 0$.
\section{The First and Second Sound in the BEC Limit}
In this section, for comparison we consider the first and second sound in the BEC limit. In the BEC limit, the system consists of bosonic molecules with mass $M=2m$ with the total number $N_{\rm{B}}=N/2$. The $s$-wave scattering length between molecule $a_{\rm{B}}$ is given in terms of the atomic scattering length $a_s$ as $a_{\rm{B}}\simeq 0.62a_s$ \cite{T_PRA93}. 
In this section, for simplicity we calculate the thermodynamic quantities and sound velocities of a dilute Bose gas within the framework of Hatrree-Fock-Bogoliubov-Popov (HFB-Popov) approximation \cite{T_HFB-popov_Griffin}. 
This means that we assume an extreme BEC limit. Solving the Gross-Pitaevskii equation and the Bogoliubov equations, within HFB-Popov approximation, we can calculate the condensate density $n_0$ and the noncondensate density $\tilde{n}$ as  
\begin{eqnarray}
\tilde{n}=\sum_{k}\frac{1}{V}\left[\frac{\epsilon_k^0+g_{\rm{B}}n_0}{E_k}f_{\rm{BE}}\left(E_k\right)+\frac{1}{2}\left(\frac{\epsilon_k^0+g_{\rm{B}}n_0}{E_k}-1\right)\right],~~n_0=n_{\rm{B}}-\tilde{n}. \label{nti}
\end{eqnarray}
where $f_{\rm{BE}}(E)=\frac{1}{\exp(\beta E)-1}$ is the Bose-Einstein distribution function and 
\begin{eqnarray}\epsilon_k^0=\frac{\hbar^2 k^2}{2M},~E_k=\sqrt{\epsilon_k^0(\epsilon_k^0+2g_{\rm{B}}n_0)}.\label{muB}\end{eqnarray} 
Equations~(\ref{nti}) and (\ref{muB}) must be calculated self-consistently. The normal fluid density is given by $n_n=\frac{\beta}{3}\sum_k k^2\frac{\partial f_{\rm{BE}}(E_k)}{\partial E_k}$ and superfluid density is given by $n_s=n_0-n_n$. 
\begin{figure}[htbp]
\centerline{\includegraphics[height=2.0in]{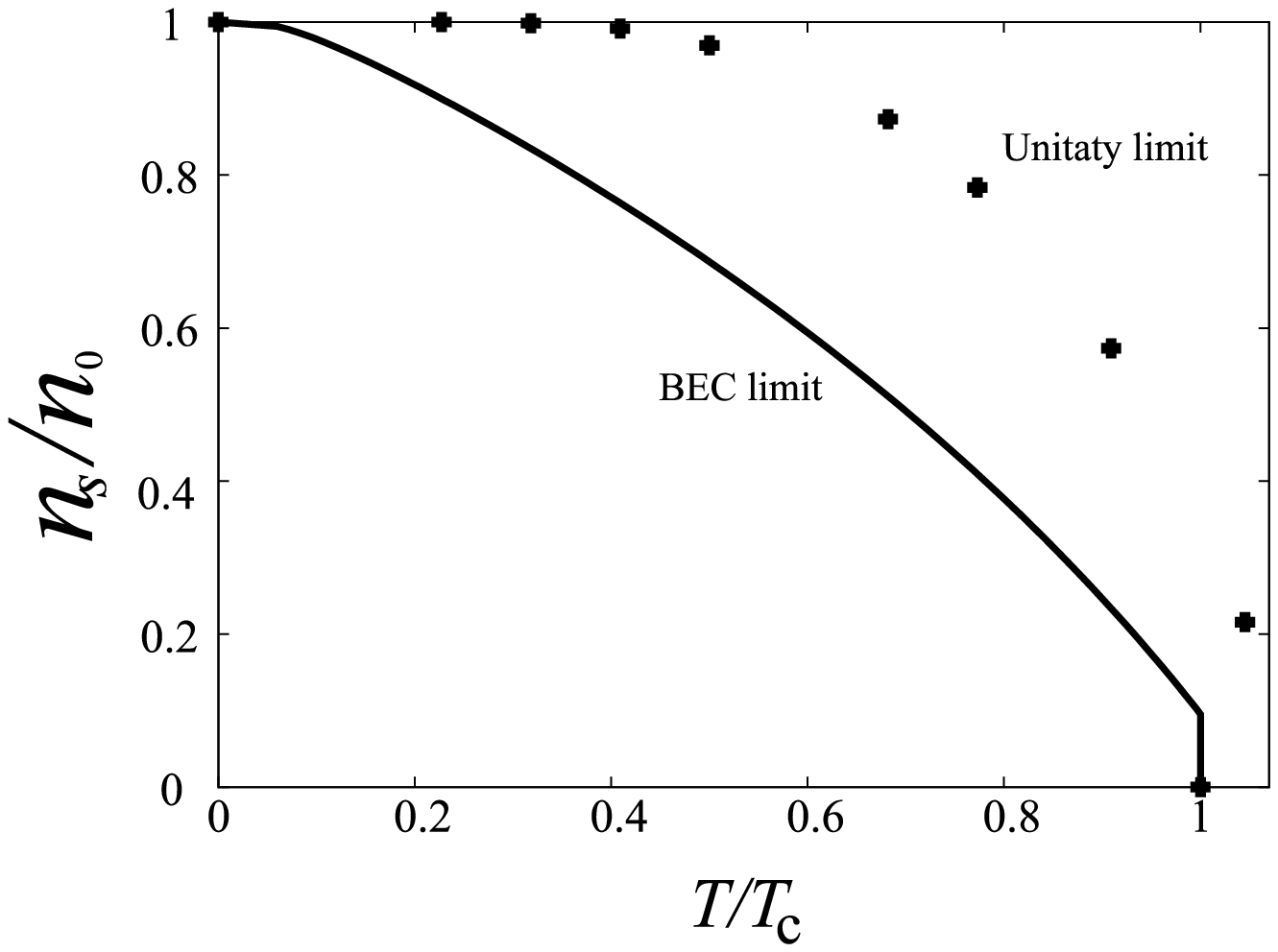}}
 \caption{Superfluid (condensation) density fraction in the BEC and unitary limit as a function of temperature.}
  \label{fig:becnc}
\end{figure}
In Fig.~\ref{fig:becnc}, we plot the temperature dependence of the superfluid density $n_s$. For comparison, we also plot $n_s$ in the unitary limit. \par
The thermodynamic functions can be calculated from the thermodynamic potential 
$P=-\frac{\Omega}{V},$ where 
\begin{eqnarray}
\Omega=-\mu n_0V +\frac{1}{2}gn_0^2V+k_{\rm{B}}T\sum_{k}\ln\left[1-e^{-\beta E_k}\right].
\end{eqnarray}
\begin{figure}[htbp]
\centerline{\includegraphics[height=2.0in]{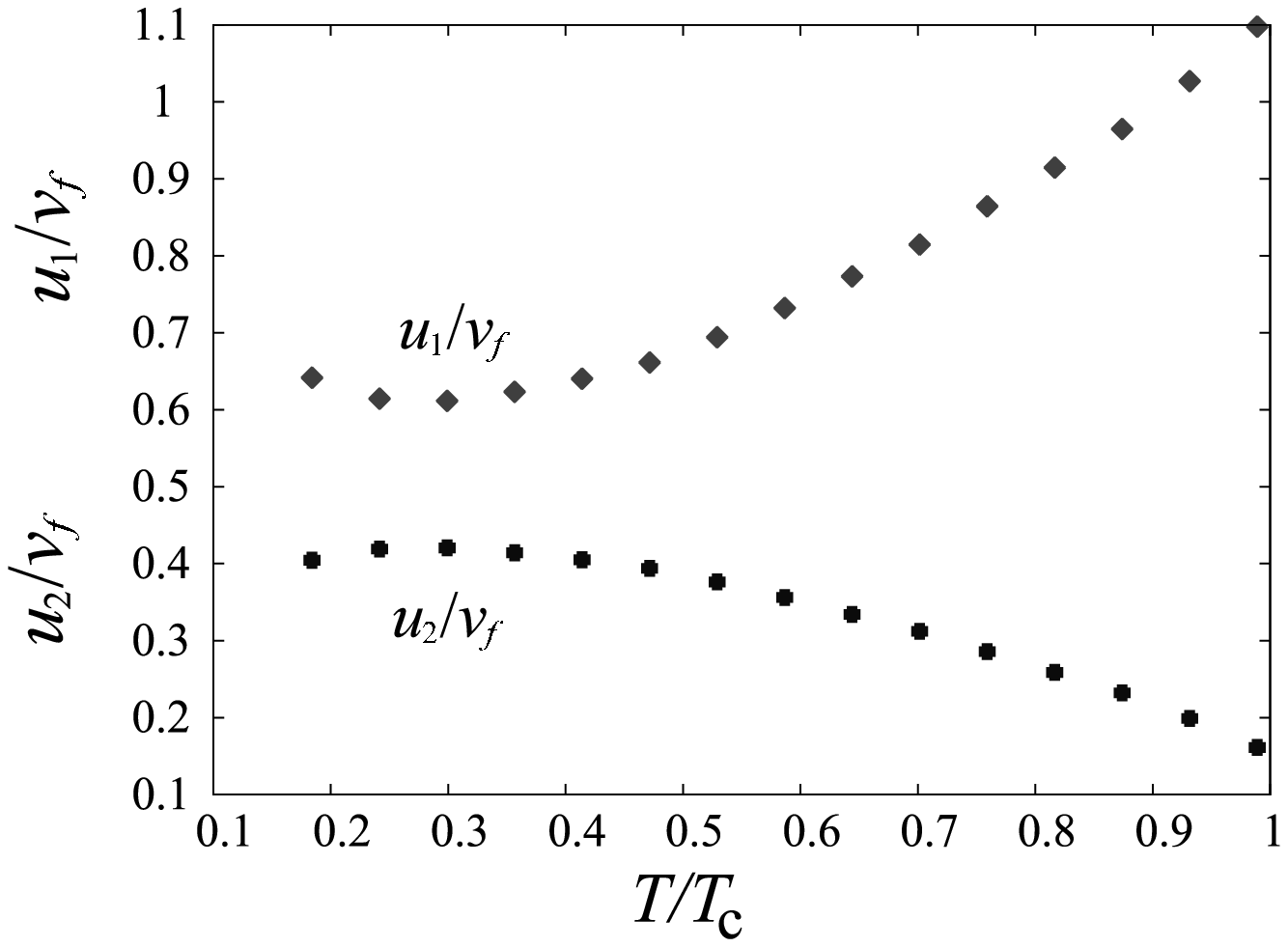}}
 \caption{The first and second sounds velocities as functions of the temperature with fixed density $gn_{\rm{B}}/k_BT_{\rm{c}}=0.15$.}
  \label{fig:becU2_U1}
\end{figure}
\begin{figure}[htbp]
\centerline{\includegraphics[height=2.0in]{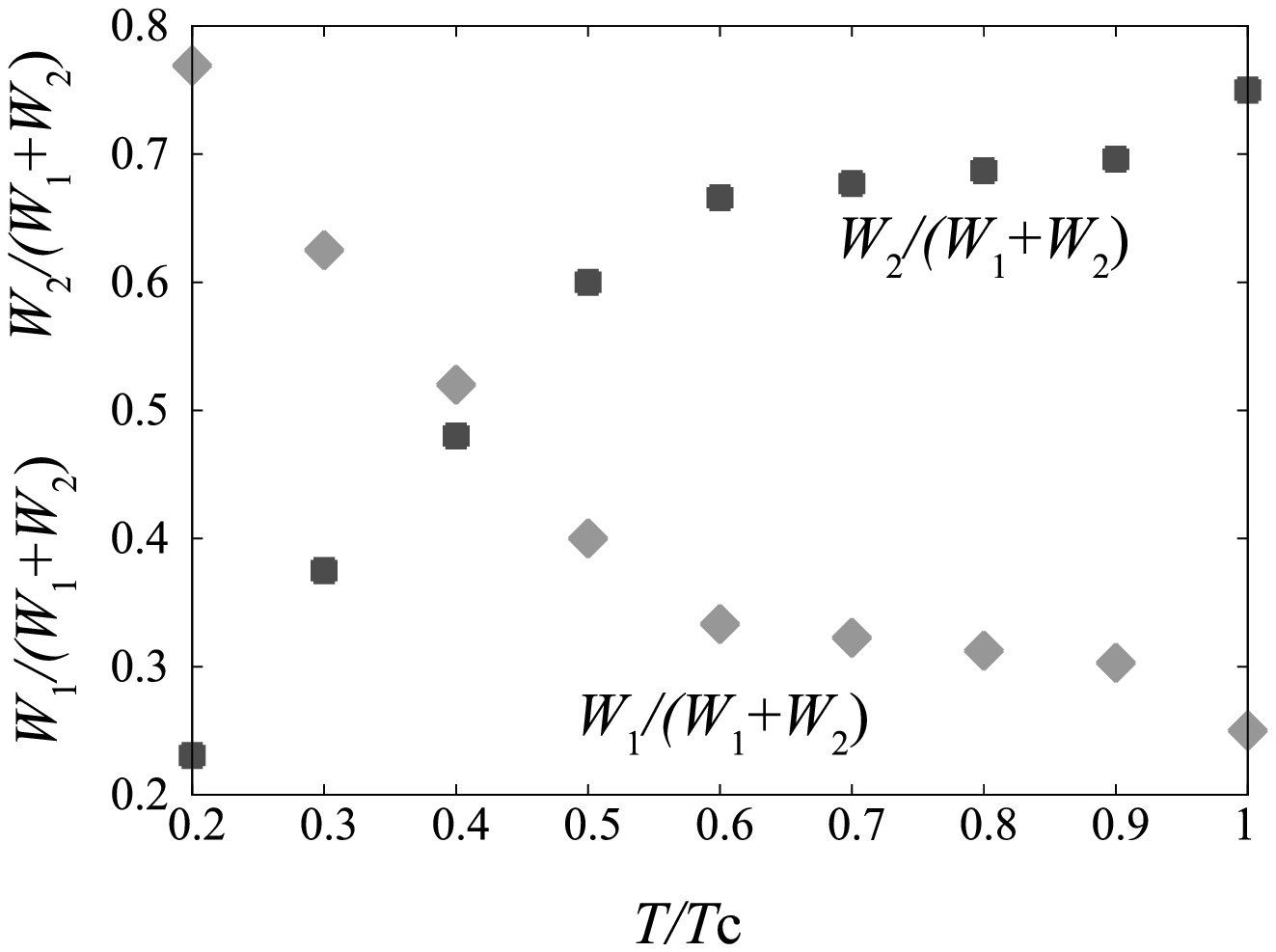}}
 \caption{$W_1/(W_1+W_2)$ and $W_2/(W_1+W_2)$ as a function of temperature with fixed density $g_{\rm{B}}n_{\rm{B}}/k_BT_{\rm{c}}=0.807$.}
  \label{fig:becW2_W1}
\end{figure}
In Fig.~\ref{fig:becU2_U1}, we plot the first and sound velocities as a function of temperature obtained from self-consistent calculation of Eqs.(\ref{nti}) and (\ref{muB}). The sound velocities are normalized by the ``Fermi velocity" defined in terms of the density and mass of bosons, i.e.  $v_f=(6\pi^2 n_{\rm B})^{1/3}/2M$. We fixed the parameters as $g_{\rm{B}}n_{\rm{B}}/k_BT_{\rm{c}}=0.807$ where $T_{\rm{c}}=\frac{2\pi \hbar^2}{m}\left(\frac{n}{2.612}\right)^{2/3}/k_B$ is the BEC transition temperature. The qualitatively similar results are obtained in Ref.~\cite{T_Gay_sound,T_Griffin_sound}, within the Hartree-Fock approximation.  
In Fig.~\ref{fig:becW2_W1}, we plot the temperature dependence of $W_1$ and $W_2$. 
When compared with Fig.~\ref{fig:W2}, the qualitative behaviors of
$W_1$ and $W_2$ in the BEC limit are quite different from those in the unitary limit, i.e. $W_2>W_1$ in most temperatures ($T>0.4T_{\rm c}$). 
This is mainly due to the difference of the temperature dependence in the superfluid density, as shown in Fig.~\ref{fig:becnc}. From Eq.~(\ref{W12}), we see that the ratio $W_1/W_2$ is determined by $v^2$, which is proportional to $\rho_{s0}/\rho_{n0}$.

\section{conclusion}
In this paper, we have discussed the propagations of the first and second sound pulses in a Fermi superfluid at unitarity. The pulse propagations are discussed in terms of the density response function obtained from Landau's two-fluid equations. In order to obtain all the thermodynamic quantities required for calculating the sound velocities and their amplitudes of the first and second sound pulses, we use the NSR-type Gaussian fluctuation theory. The results for the sound velocities are consistent with Ref.~\cite{T_Heiselberg_Unitarysound,T_condmat09050257}. 
We calculated the temperature dependence of the amplitudes of the first and second sound mode pulses, and showed that second sound pulse has an appreciable amplitude. Our results show that second sound can be excited by the pulse propagation experiment and should be observed as a separate contribution from first sound. We hope that our results will stimulate further experiment on sound pulse propagation in a strongly Fermi gas in the two-fluid hydrodynamic regime.\par 
For composition, we also calculated the temperature dependence of velocity and amplitudes of second sound pulse in the BEC limit. We showed that the qualitative behaviors of
$W_1$ and $W_2$ in the BEC limit are quite different from those in the unitary limit. This different is mainly due to the difference of the temperature dependence in the superfluid density.
\par
 Our work is based on a NSR-type Gaussian fluctuation theory \cite{T_Hu_PRA73,T_NSR,T_Taylor_PRA74,T_Fukushima_PRA75}. The NSR theory does have a problem near $T_{\rm{c}}$ near unitarity and on the BEC side of the crossover as a result of only considering Gaussian fluctuations. 
A more sophisticated theory will be required to obtain the results valid near $T_{\rm{c}}$. \section{ACKNOWLEDGMENTS} 
We thank A. Griffin for valuable comments. 
This research was supported by Academic Frontier Project (2005) of MEXT.
E. A. is supported by a Grant-in-Aid from JSPS.
\appendix
\section{Definition of $\Xi(\mathbf{q},i\nu_n)$}\label{C_1}
 \begin{eqnarray}
\Xi(\mathbf{q},i\nu_n)=\frac{1}{4}\left(\begin{array}{cc} \Pi_{11}^0+ \Pi_{22}^{0}+i( \Pi^0_{12}- \Pi^0_{21}) &  \Pi_{11}^0- \Pi_{22}^{0} \\  \Pi_{11}^0- \Pi_{22}^{0} & \Pi_{11}^0+\Pi_{22}^{0}-i( \Pi^0_{12}- \Pi^0_{21}) \end{array}\right)
\end{eqnarray},
  \begin{eqnarray}
  \Pi_{11}^0=\sum_{p}\left(1-\frac{\xi_{p+q/2}\xi_{p-q/2}-\Delta^2}{E_{p+q/2}E_{p-q/2}}\right)\frac{E_{p+q/2}-E_{p-q/2}}{(E_{p+q/2}-E_{p-q/2})^2+\nu^2_n}\nonumber \\ 
  \times \left[f_{\rm{FD}}(E_{p+q/2})-f_{\rm{FD}}(E_{p-q/2})\right]\nonumber \\ 
   -\sum_{p}\left(1+\frac{\xi_{p+q/2}\xi_{p-q/2}-\Delta^2}{E_{p+q/2}E_{p-q/2}}\right)\frac{E_{p+q/2}+E_{p-q/2}}{(E_{p+q/2}+E_{p-q/2})^2+\nu^2_n}\nonumber \\ 
  \times \left[1-f_{\rm{FD}}(E_{p+q/2})-f_{\rm{FD}}(E_{p-q/2})\right], \\ 
   \Pi_{22}^0=\sum_{p}\left(1-\frac{\xi_{p+q/2}\xi_{p-q/2}+\Delta^2}{E_{p+q/2}E_{p-q/2}}\right)\frac{E_{p+q/2}-E_{p-q/2}}{(E_{p+q/2}-E_{p-q/2})^2+\nu^2_n}\nonumber \\ 
  \times \left[f_{\rm{FD}}(E_{p+q/2})-f_{\rm{FD}}(E_{p-q/2})\right]\nonumber \\ 
   -\sum_{p}\left(1+\frac{\xi_{p+q/2}\xi_{p-q/2}+\Delta^2}{E_{p+q/2}E_{p-q/2}}\right)\frac{E_{p+q/2}+E_{p-q/2}}{(E_{p+q/2}+E_{p-q/2})^2+\nu^2_n}\nonumber \\ 
  \times \left[1-f_{\rm{FD}}(E_{p+q/2})-f_{\rm{FD}}(E_{p-q/2})\right], \\
     \Pi_{22}^0=\sum_{p}\left(\frac{\xi_{p+q/2}}{E_{p+q/2}}-\frac{\xi_{p-q/2}}{E_{p-q/2}}\right)\frac{\nu_n}{(E_{p+q/2}-E_{p-q/2})^2+\nu^2_n}\nonumber \\
     \times \left[f_{\rm{FD}}(E_{p+q/2})-f_{\rm{FD}}(E_{p-q/2})\right] \nonumber \\
     -\sum_{p}\left(\frac{\xi_{p+q/2}}{E_{p+q/2}}+\frac{\xi_{p-q/2}}{E_{p-q/2}}\right)\frac{\nu_n}{(E_{p+q/2}+E_{p-q/2})^2+\nu^2_n}\nonumber \\
     \times \left[1-f_{\rm{FD}}(E_{p+q/2})-f_{\rm{FD}}(E_{p-q/2})\right]\nonumber \\
     =-\Pi_{21}^0.
\end{eqnarray}

\begin{thebibliography}{29}
\expandafter\ifx\csname natexlab\endcsname\relax\def\natexlab#1{#1}\fi
\expandafter\ifx\csname bibnamefont\endcsname\relax
  \def\bibnamefont#1{#1}\fi
\expandafter\ifx\csname bibfnamefont\endcsname\relax
  \def\bibfnamefont#1{#1}\fi
\expandafter\ifx\csname citenamefont\endcsname\relax
  \def\citenamefont#1{#1}\fi
\expandafter\ifx\csname url\endcsname\relax
  \def\url#1{\texttt{#1}}\fi
\expandafter\ifx\csname urlprefix\endcsname\relax\def\urlprefix{URL }\fi
\providecommand{\bibinfo}[2]{#2}
\providecommand{\eprint}[2][]{\url{#2}}

\bibitem[{\citenamefont{Landau}(1941)}]{Hydro_Landau}
\bibinfo{author}{\bibfnamefont{L.~D.} \bibnamefont{Landau}},
  \bibinfo{journal}{J. Phys. (USSR)} \textbf{\bibinfo{volume}{71}},
  \bibinfo{pages}{5} (\bibinfo{year}{1941}).

\bibitem[{\citenamefont{Kinast et~al.}(2004)\citenamefont{Kinast, Hemmer, Gehm,
  Turlapov, and Thomas}}]{E_Kinast_PRL92}
\bibinfo{author}{\bibfnamefont{J.}~\bibnamefont{Kinast}},
  \bibinfo{author}{\bibfnamefont{S.~L.} \bibnamefont{Hemmer}},
  \bibinfo{author}{\bibfnamefont{M.~E.} \bibnamefont{Gehm}},
  \bibinfo{author}{\bibfnamefont{A.}~\bibnamefont{Turlapov}}, \bibnamefont{and}
  \bibinfo{author}{\bibfnamefont{J.~E.} \bibnamefont{Thomas}},
  \bibinfo{journal}{Phys. Rev. Lett.} \textbf{\bibinfo{volume}{92}},
  \bibinfo{pages}{150402} (\bibinfo{year}{2004}).

\bibitem[{\citenamefont{Bartenstein et~al.}(2004)\citenamefont{Bartenstein,
  Altmeyer, Riedl, Jochim, Chin, Denschlag, and Grimm}}]{E_Bartenstein_PRL92}
\bibinfo{author}{\bibfnamefont{M.}~\bibnamefont{Bartenstein}},
  \bibinfo{author}{\bibfnamefont{A.}~\bibnamefont{Altmeyer}},
  \bibinfo{author}{\bibfnamefont{S.}~\bibnamefont{Riedl}},
  \bibinfo{author}{\bibfnamefont{S.}~\bibnamefont{Jochim}},
  \bibinfo{author}{\bibfnamefont{C.}~\bibnamefont{Chin}},
  \bibinfo{author}{\bibfnamefont{J.~H.} \bibnamefont{Denschlag}},
  \bibnamefont{and} \bibinfo{author}{\bibfnamefont{R.}~\bibnamefont{Grimm}},
  \bibinfo{journal}{Phys. Rev. Lett} \textbf{\bibinfo{volume}{92}},
  \bibinfo{pages}{203201} (\bibinfo{year}{2004}).

\bibitem[{\citenamefont{Joseph et~al.}(2007)\citenamefont{Joseph, Clancy,
  L.~Luo, Turlapov, and Thomas}}]{E_Joseph_PRL98}
\bibinfo{author}{\bibfnamefont{J.}~\bibnamefont{Joseph}},
  \bibinfo{author}{\bibfnamefont{B.}~\bibnamefont{Clancy}},
  \bibinfo{author}{\bibfnamefont{L.}~ \bibnamefont{Luo}},
   \bibinfo{author}{\bibfnamefont{J.}~ \bibnamefont{Kinast}},
  \bibinfo{author}{\bibfnamefont{A.}~\bibnamefont{Turlapov}}, \bibnamefont{and}
  \bibinfo{author}{\bibfnamefont{J.~E.} \bibnamefont{Thomas}},
  \bibinfo{journal}{Phys. Rev. Lett.} \textbf{\bibinfo{volume}{98}},
  \bibinfo{pages}{170401} (\bibinfo{year}{2007}).

\bibitem[{\citenamefont{Taylor and Griffin}(2005)}]{T_Taylo_sound}
\bibinfo{author}{\bibfnamefont{E.}~\bibnamefont{Taylor}} \bibnamefont{and}
  \bibinfo{author}{\bibfnamefont{A.}~\bibnamefont{Griffin}},
  \bibinfo{journal}{Phys. Rev. A} \textbf{\bibinfo{volume}{72}},
  \bibinfo{pages}{053630} (\bibinfo{year}{2005}).

\bibitem[{\citenamefont{Heiselberg}(2006)}]{T_Heiselberg_Unitarysound}
\bibinfo{author}{\bibfnamefont{H.}~\bibnamefont{Heiselberg}},
  \bibinfo{journal}{Phys. Rev. A} \textbf{\bibinfo{volume}{73}},
  \bibinfo{pages}{013607} (\bibinfo{year}{2006}).

\bibitem[{\citenamefont{Fukushima et~al.}(2007)\citenamefont{Fukushima, Ohashi,
  Taylor, and Griffin}}]{T_Fukushima_PRA75}
\bibinfo{author}{\bibfnamefont{N.}~\bibnamefont{Fukushima}},
  \bibinfo{author}{\bibfnamefont{Y.}~\bibnamefont{Ohashi}},
  \bibinfo{author}{\bibfnamefont{E.}~\bibnamefont{Taylor}}, \bibnamefont{and}
  \bibinfo{author}{\bibfnamefont{A.}~\bibnamefont{Griffin}},
  \bibinfo{journal}{Phys. Rev. A} \textbf{\bibinfo{volume}{75}},
  \bibinfo{pages}{033609} (\bibinfo{year}{2007}).

\bibitem[{\citenamefont{Taylor et~al.}(2008)\citenamefont{Taylor, Hu, Liu, and
  Griffin}}]{T_Taylor_PRA77}
\bibinfo{author}{\bibfnamefont{E.}~\bibnamefont{Taylor}},
  \bibinfo{author}{\bibfnamefont{H.}~\bibnamefont{Hu}},
  \bibinfo{author}{\bibfnamefont{X.-J.} \bibnamefont{Liu}}, \bibnamefont{and}
  \bibinfo{author}{\bibfnamefont{A.}~\bibnamefont{Griffin}},
  \bibinfo{journal}{Phys. Rev. A} \textbf{\bibinfo{volume}{77}},
  \bibinfo{pages}{033608} (\bibinfo{year}{2008}).

\bibitem[{\citenamefont{Taylor et~al.}(2007)\citenamefont{Taylor, Hu, Liu,
  Pitaevskii, and Griffin}}]{T_condmat07090698}
\bibinfo{author}{\bibfnamefont{E.}~\bibnamefont{Taylor}},
  \bibinfo{author}{\bibfnamefont{H.}~\bibnamefont{Hu}},
  \bibinfo{author}{\bibfnamefont{X.-J.} \bibnamefont{Liu}},
  \bibinfo{author}{\bibfnamefont{L.~P.} \bibnamefont{Pitaevskii}},
  \bibnamefont{and} \bibinfo{author}{\bibfnamefont{A.}~\bibnamefont{Griffin}},
  \bibinfo{journal}{arXiv:0709.0698v2}  (\bibinfo{year}{2007}).

\bibitem[{\citenamefont{Taylor et~al.}(2009)\citenamefont{Taylor, Hu, Liu,
  Pitaevskii, Griffin, and Stringari}}]{T_condmat09050257}
\bibinfo{author}{\bibfnamefont{E.}~\bibnamefont{Taylor}},
  \bibinfo{author}{\bibfnamefont{H.}~\bibnamefont{Hu}},
  \bibinfo{author}{\bibfnamefont{X.-J.} \bibnamefont{Liu}},
  \bibinfo{author}{\bibfnamefont{L.~P.} \bibnamefont{Pitaevskii}},
  \bibinfo{author}{\bibfnamefont{A.}~\bibnamefont{Griffin}}, \bibnamefont{and}
  \bibinfo{author}{\bibfnamefont{S.}~\bibnamefont{Stringari}},
  \bibinfo{journal}{arXiv:0905.0257v1}  (\bibinfo{year}{2009}).

\bibitem[{\citenamefont{He et~al.}(2007)\citenamefont{He, Chen, Chien, and
  Levin}}]{T_Yan_sound}
\bibinfo{author}{\bibfnamefont{Y.}~\bibnamefont{He}},
  \bibinfo{author}{\bibfnamefont{Q.}~\bibnamefont{Chen}},
  \bibinfo{author}{\bibfnamefont{C.-C.} \bibnamefont{Chien}}, \bibnamefont{and}
  \bibinfo{author}{\bibfnamefont{K.}~\bibnamefont{Levin}},
  \bibinfo{journal}{Phys. Rev. A} \textbf{\bibinfo{volume}{76}},
  \bibinfo{pages}{051602(R)} (\bibinfo{year}{2007}).

\bibitem[{\citenamefont{Andrews et~al.}(1997)\citenamefont{Andrews, Kurn,
  Miesner, Durfee, Townsend, Inouye, and Ketterle}}]{E_Andrews_PRL79}
\bibinfo{author}{\bibfnamefont{M.~R.} \bibnamefont{Andrews}},
  \bibinfo{author}{\bibfnamefont{D.~M.} \bibnamefont{Kurn}},
  \bibinfo{author}{\bibfnamefont{H.-J.} \bibnamefont{Miesner}},
  \bibinfo{author}{\bibfnamefont{D.~S.} \bibnamefont{Durfee}},
  \bibinfo{author}{\bibfnamefont{C.~G.} \bibnamefont{Townsend}},
  \bibinfo{author}{\bibfnamefont{S.}~\bibnamefont{Inouye}}, \bibnamefont{and}
  \bibinfo{author}{\bibfnamefont{W.}~\bibnamefont{Ketterle}},
  \bibinfo{journal}{Phys. Rev. Lett.} \textbf{\bibinfo{volume}{79}},
  \bibinfo{pages}{553} (\bibinfo{year}{1997}).

\bibitem[{\citenamefont{Griffin et~al.}(1997)\citenamefont{Griffin, Wu, and
  Stringari}}]{T_PRA78}
\bibinfo{author}{\bibfnamefont{A.}~\bibnamefont{Griffin}},
  \bibinfo{author}{\bibfnamefont{W.-C.} \bibnamefont{Wu}}, \bibnamefont{and}
  \bibinfo{author}{\bibfnamefont{S.}~\bibnamefont{Stringari}},
  \bibinfo{journal}{Phys. Rev. Lett.} \textbf{\bibinfo{volume}{78}},
  \bibinfo{pages}{1838} (\bibinfo{year}{1997}).

\bibitem[{\citenamefont{Zaremba}(1998)}]{T_PRA57}
\bibinfo{author}{\bibfnamefont{E.}~\bibnamefont{Zaremba}},
  \bibinfo{journal}{Phys. Rev. A} \textbf{\bibinfo{volume}{57}},
  \bibinfo{pages}{518} (\bibinfo{year}{1998}).

\bibitem[{\citenamefont{Nikuni and Griffin}(1998)}]{T_Nikuni_}
\bibinfo{author}{\bibfnamefont{T.}~\bibnamefont{Nikuni}} \bibnamefont{and}
  \bibinfo{author}{\bibfnamefont{A.}~\bibnamefont{Griffin}},
  \bibinfo{journal}{Phys. Rev. A} \textbf{\bibinfo{volume}{58}},
  \bibinfo{pages}{4044} (\bibinfo{year}{1998}).

\bibitem[{\citenamefont{Capuzzi et~al.}(2006)\citenamefont{Capuzzi, Vignolo,
  Federici, and Tosi}}]{T_C}
\bibinfo{author}{\bibfnamefont{P.}~\bibnamefont{Capuzzi}},
  \bibinfo{author}{\bibfnamefont{P.}~\bibnamefont{Vignolo}},
  \bibinfo{author}{\bibfnamefont{F.}~\bibnamefont{Federici}}, \bibnamefont{and}
  \bibinfo{author}{\bibfnamefont{M.~P.} \bibnamefont{Tosi}},
  \bibinfo{journal}{Phys. Rev. A} \textbf{\bibinfo{volume}{73}},
  \bibinfo{pages}{021603(R)} (\bibinfo{year}{2006}).

\bibitem[{\citenamefont{Pethich and Smith}(2002)}]{B_Pethich_Smith}
\bibinfo{author}{\bibfnamefont{C.~J.} \bibnamefont{Pethich}} \bibnamefont{and}
  \bibinfo{author}{\bibfnamefont{H.}~\bibnamefont{Smith}},
  \emph{\bibinfo{title}{Bose-Einstein Condensation in Dilute Bose Gases}}
  (\bibinfo{publisher}{UNIVERSITY PRESS CAMBRIDGE}, \bibinfo{year}{2002}).

\bibitem[{\citenamefont{Griffin et~al.}(2009)\citenamefont{Griffin, Nikun, and
  Zaremba}}]{B_GNZ}
\bibinfo{author}{\bibfnamefont{A.}~\bibnamefont{Griffin}},
  \bibinfo{author}{\bibfnamefont{T.}~\bibnamefont{Nikun}}, \bibnamefont{and}
  \bibinfo{author}{\bibfnamefont{E.}~\bibnamefont{Zaremba}},
  \emph{\bibinfo{title}{Bose-Condensed Gases at Finite Temperatures}}
  (\bibinfo{publisher}{UNIVERSITY PRESS CAMBRIDGE}, \bibinfo{year}{2009}).

\bibitem[{\citenamefont{Ginzbrug}(1943)}]{Hydro_Ginz}
\bibinfo{author}{\bibfnamefont{V.~L.} \bibnamefont{Ginzbrug}},
  \bibinfo{journal}{J. Expt. Theor. Phys. (USSR)}
  \textbf{\bibinfo{volume}{243}}, \bibinfo{pages}{13} (\bibinfo{year}{1943}).

\bibitem[{\citenamefont{Hohenberg and Martin}(1964)}]{T_Hohenberg_PRL12}
\bibinfo{author}{\bibfnamefont{P.~C.} \bibnamefont{Hohenberg}}
  \bibnamefont{and} \bibinfo{author}{\bibfnamefont{P.~C.}
  \bibnamefont{Martin}}, \bibinfo{journal}{Phys. Rev. Lett.}
  \textbf{\bibinfo{volume}{12}}, \bibinfo{pages}{69} (\bibinfo{year}{1964}).

\bibitem[{\citenamefont{Gay and Griffin}(1985)}]{T_Gay_sound}
\bibinfo{author}{\bibfnamefont{C.}~\bibnamefont{Gay}} \bibnamefont{and}
  \bibinfo{author}{\bibfnamefont{A.}~\bibnamefont{Griffin}},
  \bibinfo{journal}{J. Low. Temp. Phys.} \textbf{\bibinfo{volume}{58}},
  \bibinfo{pages}{479} (\bibinfo{year}{1985}).

\bibitem[{\citenamefont{Nozi$\grave{e}$res and Schmitt-Rink}(1985)}]{T_NSR}
\bibinfo{author}{\bibfnamefont{P.}~\bibnamefont{Nozi$\grave{e}$res}}
  \bibnamefont{and}
  \bibinfo{author}{\bibfnamefont{S.}~\bibnamefont{Schmitt-Rink}},
  \bibinfo{journal}{J. Low. Temp. Phys.} \textbf{\bibinfo{volume}{59}},
  \bibinfo{pages}{195} (\bibinfo{year}{1985}).

\bibitem[{\citenamefont{Engelbrecht et~al.}(1997)\citenamefont{Engelbrecht,
  Randeria, and de~Melo}}]{T_Engelbrecht_PRB55}
\bibinfo{author}{\bibfnamefont{J.~R.} \bibnamefont{Engelbrecht}},
  \bibinfo{author}{\bibfnamefont{M.}~\bibnamefont{Randeria}}, \bibnamefont{and}
  \bibinfo{author}{\bibfnamefont{C.~A.~R.} \bibnamefont{Sa~ de~Melo}},
  \bibinfo{journal}{Phys. Rev. B} \textbf{\bibinfo{volume}{55}},
  \bibinfo{pages}{15153  } (\bibinfo{year}{1997}).

\bibitem[{\citenamefont{Ohashi and Griffin}(2003)}]{T_Ohashi_PRA}
\bibinfo{author}{\bibfnamefont{Y.}~\bibnamefont{Ohashi}} \bibnamefont{and}
  \bibinfo{author}{\bibfnamefont{A.}~\bibnamefont{Griffin}},
  \bibinfo{journal}{Phys. Rev. A} \textbf{\bibinfo{volume}{67}},
  \bibinfo{pages}{063612} (\bibinfo{year}{2003}).

\bibitem[{\citenamefont{Taylor et~al.}(2006)\citenamefont{Taylor, Griffin,
  Fukushima, and Ohashi}}]{T_Taylor_PRA74}
\bibinfo{author}{\bibfnamefont{E.}~\bibnamefont{Taylor}},
  \bibinfo{author}{\bibfnamefont{A.}~\bibnamefont{Griffin}},
  \bibinfo{author}{\bibfnamefont{N.}~\bibnamefont{Fukushima}},
  \bibnamefont{and} \bibinfo{author}{\bibfnamefont{Y.}~\bibnamefont{Ohashi}},
  \bibinfo{journal}{Phys. Rev. A} \textbf{\bibinfo{volume}{74}},
  \bibinfo{pages}{063626} (\bibinfo{year}{2006}).

\bibitem[{\citenamefont{Hu et~al.}(2006)\citenamefont{Hu, Liu, and
  Drummond}}]{T_Hu_PRA73}
\bibinfo{author}{\bibfnamefont{H.}~\bibnamefont{Hu}},
  \bibinfo{author}{\bibfnamefont{X.-J.} \bibnamefont{Liu}}, \bibnamefont{and}
  \bibinfo{author}{\bibfnamefont{P.~D.} \bibnamefont{Drummond}},
  \bibinfo{journal}{Phys. Rev. A} \textbf{\bibinfo{volume}{73}},
  \bibinfo{pages}{023617  } (\bibinfo{year}{2006}).

\bibitem[{\citenamefont{Petrov et~al.}(2004)\citenamefont{Petrov, Salomon, and
  Shlyapnikov}}]{T_PRA93}
\bibinfo{author}{\bibfnamefont{D.~S.} \bibnamefont{Petrov}},
  \bibinfo{author}{\bibfnamefont{C.}~\bibnamefont{Salomon}}, \bibnamefont{and}
  \bibinfo{author}{\bibfnamefont{G.~V.} \bibnamefont{Shlyapnikov}},
  \bibinfo{journal}{Phys. Rev. Lett.} \textbf{\bibinfo{volume}{93}},
  \bibinfo{pages}{090404} (\bibinfo{year}{2004}).

\bibitem[{\citenamefont{Griffin}(1996)}]{T_HFB-popov_Griffin}
\bibinfo{author}{\bibfnamefont{A.}~\bibnamefont{Griffin}},
  \bibinfo{journal}{Phys. Rev. B} \textbf{\bibinfo{volume}{53}},
  \bibinfo{pages}{9341} (\bibinfo{year}{1996}).

\bibitem[{\citenamefont{Griffin and Zaremba}(1997)}]{T_Griffin_sound}
\bibinfo{author}{\bibfnamefont{A.}~\bibnamefont{Griffin}} \bibnamefont{and}
  \bibinfo{author}{\bibfnamefont{E.}~\bibnamefont{Zaremba}},
  \bibinfo{journal}{Phys. Rev. A} \textbf{\bibinfo{volume}{56}},
  \bibinfo{pages}{4839} (\bibinfo{year}{1997}).

\end{thebibliography}

\end{document}